\documentclass[a4paper, oneside, 12pt, onecolumn]{article}
\usepackage{settings}

\title{VENu: The Virtual Environment for Neutrinos} % Article title
\author{%
\textsc{Marco Del Tutto}\thanks{\mail{marco.deltutto@physics.ox.ac.uk}} \\ % Your name
\normalsize Department of Physics, University of Oxford,\\ 
\normalsize Oxford OX1 3RH, United Kingdom \\[1ex] % Your institution
\normalsize Representing the \uB Collaboration
}
\date{\today} % Leave empty to omit a date
\renewcommand{\maketitlehookd}{%
\begin{abstract}
The Virtual Environment for Neutrinos (VENu) is a virtual reality-based visualisation of the MicroBooNE detector. MicroBooNE is a liquid-argon-based neutrino experiment, which is currently operating in Fermilab’s Booster neutrino beam. 
The new VENu smartphone app provides informative explanations about neutrinos and uses real MicroBooNE neutrino data that can be visualised inside a virtual representation of the MicroBooNE detector. Available for both iOS and Android, the VENu app can be downloaded for free from the Apple and Google marketplaces. The app enables users to immerse themselves inside the MicroBooNE particle detector and to see particle tracks inside. This can be done in Virtual Reality mode, where the users can pair their smartphone with any consumer virtual reality headset and see the detector in 3D. To encourage learning in a fun environment, a game is also available, guiding users to learn about neutrinos and how to detect them. They can also try to ``catch'' neutrinos themselves in 3D mode. The app is currently being pursued for a QuarkNet neutrino master class and outreach events at several universities and labs worldwide. 
\end{abstract}
}
\begin{document}

\maketitle

%\listoftodos

%\tableofcontents

\begin{center}
Talk presented at the APS Division of Particles and Fields Meeting (DPF 2017), July 31-August 4, 2017, Fermilab. C170731
\end{center}

\newpage

\section{Introduction}

Neutrino detection is entering a new era with a number of high precision neutrino detectors, specifically Liquid Argon Time Projection Chambers (LArTPCs). 
%LArTPCs are a new technology for neutrino detection. 
At MicroBooNE we have developed a new way to show events recorded by LArTPCs, using an interactive Virtual Reality experience. These proceedings describe \emph{VENu} (Virtual Environment for Neutrinos): a new event display to show LArTPCs events.

The Micro Booster Neutrino Experiment (\uB), is the first large ($86$ tons) LArTPC to operate in the United States, \cite{tdr}. LArTPC detectors combine fine-grained topology with total absorption calorimetry to provide excellent signal efficiency and background rejection. \uB is currently taking data using the Fermilab's Booster Neutrino Beam (BNB).

The MicroBooNE experiment combines physics goals of short-baseline oscillations and neutrino cross-section measurements with development goals to inform larger scale construction of LArTPCs for the long-baseline neutrino program. 
%MicroBooNE's principal physics goal is to address the long standing unsolved puzzles generated by short-baseline neutrino oscillation results from past experiments. 

\section{The VENu Mobile App}

\begin{figure}[]
\centering
\includegraphics[width=0.98\textwidth]{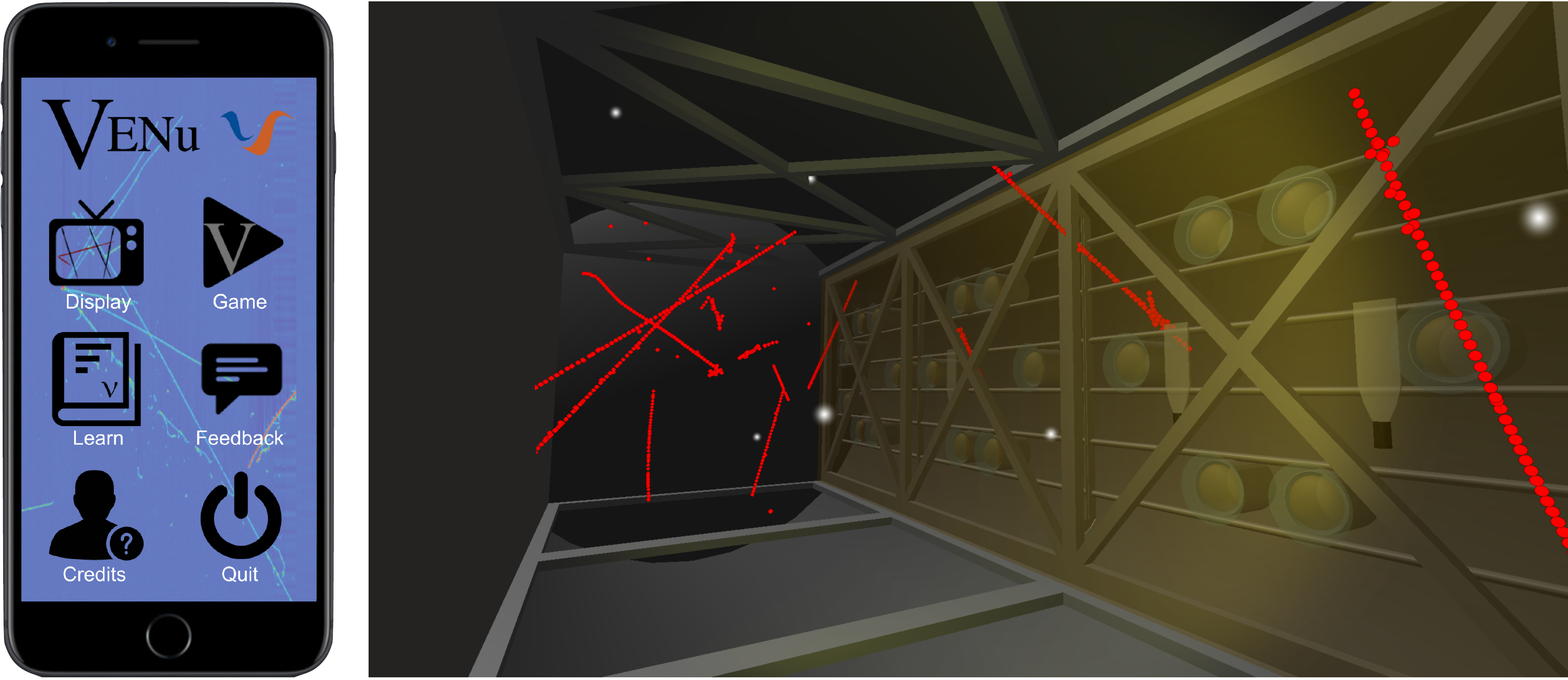}
\caption{VENu event display. Red dots are 3D reconstructed points of the particle along its path taken from a MicroBooNE data event (run 5975, event 4262). While many cosmic rays are visible, there is also a neutrino candidate event at the end of the detector, where a candidate muon track, a proton and two photons coming from a candidate $\pi^0$ decay are visible. The circles on the right side of the detector represent the MicroBooNE photomultiplier system. The white dots are an artistic representations of the neutrinos from the beam.}
\captionsetup{format=hang,labelfont={sf,bf}}
\label{fig:iphone_evd}
\end{figure}

VENu is a mobile app available for both iOS and Android devices \cite{ios_download} \cite{android_download}. The app contains several features, the main one being a 3D rendered event display, as can be seen in Figure \ref{fig:iphone_evd}. The display shows the inside of the \uB TPC. The red dots are 3D reconstructed points, that basically represent the particle trajectories when traversing the detector. The white dots are an artistic representation of the neutrinos that are constantly bombarding the detector to give an idea of where neutrinos are coming from while the display is running.

The app also provides informative explanations about neutrinos and uses real MicroBooNE neutrino data. It enables users to immerse themselves inside the MicroBooNE particle detector and to see the particles that interacted in it. This can also be done in Virtual Reality (VR) mode where the users can pair their smartphone with any consumer virtual reality headset and see the detector in 3D, such as the one shown in Figure \ref{fig:venu12}. 

The app is available to download for free from the Apple App Store \cite{ios_download} and the Google Play Store \cite{android_download} and is currently being used for physics outreach events. 

To encourage learning in a fun environment, a game is also available. The game guides users in understanding what neutrinos are and what are the techniques used to detect them. As described in more details below, users can also try to catch neutrinos themselves in 3D mode by being asked to identify the neutrino interactions in the detector.

The app can also connect to social media platforms, allowing users to share their progress and achievement when playing the game.

%+++++++++FIGURE+++++++++
%++++++++++++++++++++++++

\section{Motivations}

The VENu app was created with the intention of providing a tool for physics outreach and of offering a tool for neutrino physicists to interact with the public while describing their research. 
The idea of developing an app came out as one of the best ways of reaching a large number of people without having the constant support of a scientist to explain what the research is about. Also the costs of creating and using an app are modest and the impact can be very high.

Nowadays 3D visualisation using Virtual Reality is being widely used and at MicroBooNE we realised quite early on the power and impact that this can have. We also wanted an innovative way to show the experiment to the public. This is why the app has a Virtual Reality version of the event display, that can be used combining the smartphone with a Google Cardboard \cite{cardboard} headset (Figure \ref{fig:venu12}).

A series of learning sections about neutrinos, detector, etc. have also been added to make the app self explanatory. The available game allows users to connect the various learning sections. The game tutorial guides the users into several steps from which they can learn more about neutrino physics.

The educational game was motivated by the fact that young people will use it to hunt neutrinos and to learn more about them in a fun environment.\\

%+++++++++FIGURE+++++++++
\begin{figure}[t]
%\begin{adjustwidth}{-2cm}{-2cm}
\centering
\subfloat[][An example of a tutorial section describing the MicroBooNE experiment.]
   {\includegraphics[width=.225\textwidth]{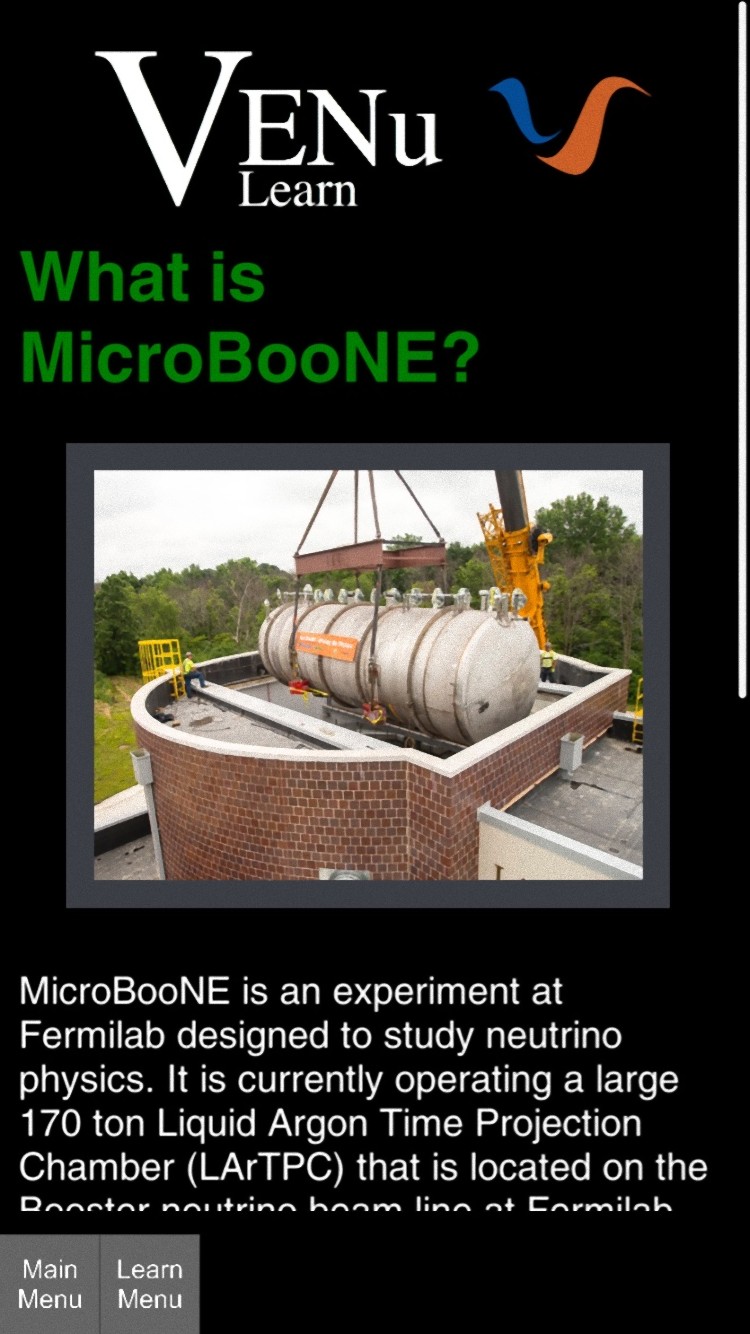}
   \label{fig:venu_learn_microboone}} \quad
\subfloat[][The VENu event display in Virtual Reality mode. Placing the smartphone in a VR viewer allows the user to have a full 3D experience.]
   {\includegraphics[width=.71\textwidth]{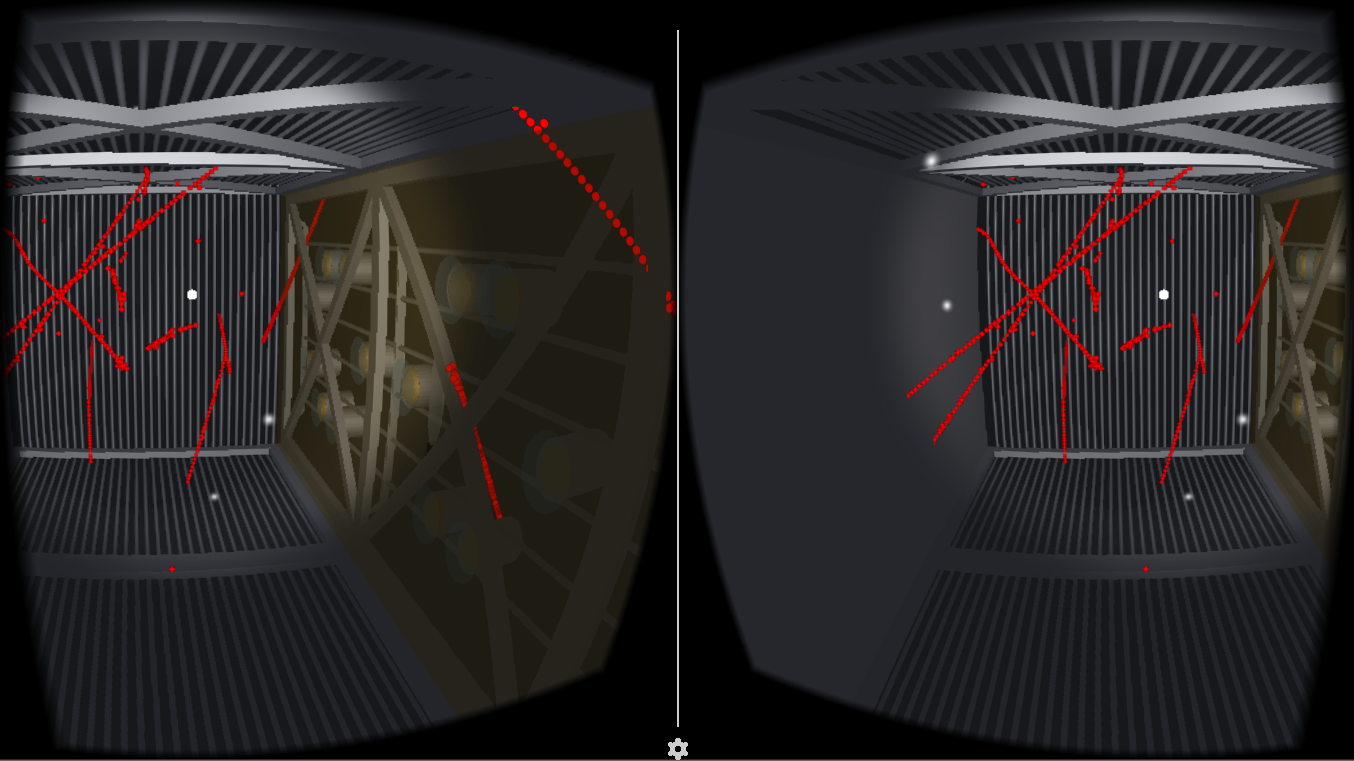}
   \label{fig:cardboardScreen}} \\
\caption{}
\label{fig:3d}
%\end{adjustwidth}
\end{figure}
%++++++++++++++++++++++++

\section{Development}

The project involved the following phases: the design, development and launch of the final app on the Apple and Google marketplaces.
VENu is built and rendered in a 3D environment using the game engine Unity \cite{unity}. Unity is an open source video game engine with very powerful multi-platform portability. All 3D models and scripts are assembled inside of Unity, and all code is written in \texttt{JavaScript} (\texttt{UnityScript}) or \texttt{C\#}.

\subsection{Rendering of the MicroBooNE detector}

The MicroBooNE detector geometry has been rendered using the Blender software \cite{blender}. Blender is an open source 3D modelling software, that imports easily into Unity. The arbitrary units/dimensions in Blender port directly over to Unity. The cryostat, TPC, and locations of PMT's are dimensionally correct. Everything else (i.e. feedthroughs) was depicted approximately \cite{geom}.

\subsection{Rendering of the MicroBooNE data}

Data from the MicroBooNE detector are first processed using the LArSoft software \cite{larsoft}. LArSoft is a suite of tools designed for use in liquid argon TPCs, providing reconstruction, particle identification, electromagnetic shower reconstruction and optical reconstruction: all the tools required for a complete picture of neutrino interactions in a liquid argon TPC.

The data are then processed in a simplified \texttt{json} (JavaScript Object Notation) format. This is done using tools already available in the experiment, particularly thanks to the already existing Argo event display \cite{argo}.

A Unity script, that runs exclusively in the Unity Editor, takes the \texttt{json} files as input and it makes \texttt{prefabs} objects. \texttt{prefabs} in Unity are assets that allow to store a game object (like a particle trajectory). These objects are basically lines and points used to display particle's trajectories or just 3D reconstructed points of a particle along its path.

These \texttt{prefabs} are then saved and stored in the app, and will be called at runtime when the user wants to display them. This saves time and gives a better user experience, as the \texttt{json} files are usually too large to be parsed at runtime.

\section{Deployment}

\subsection{Testing}

A beta version of the app was tested by MicroBooNE collaborators. All the needed modifications were implemented based on the feedback received. 
A very important testing was done by visiting Matthew Arnold School in Oxford (UK) and showing the app to about 30 students (12-16 years old). The overall impression the students had was really good but much useful
input was given. For example: the need to introduce an explanation describing the commands to move inside the detector; to clarify more in the learning sections and add different levels in the game.

\subsection{Launch}

The app was launched at the University of Oxford during the Stargazing event \cite{stargazing} in the Physics department on January 28, 2017. It was targeted to families. Being an educational app, it can be downloaded on the kids' or one of the parents' phones. There were over 1,000 visitors at the event.

\begin{figure}[t]
%\begin{adjustwidth}{-2cm}{-2cm}
\centering
\subfloat[][VENu branded Google Cardboards.]
   {\includegraphics[width=.3\textwidth]{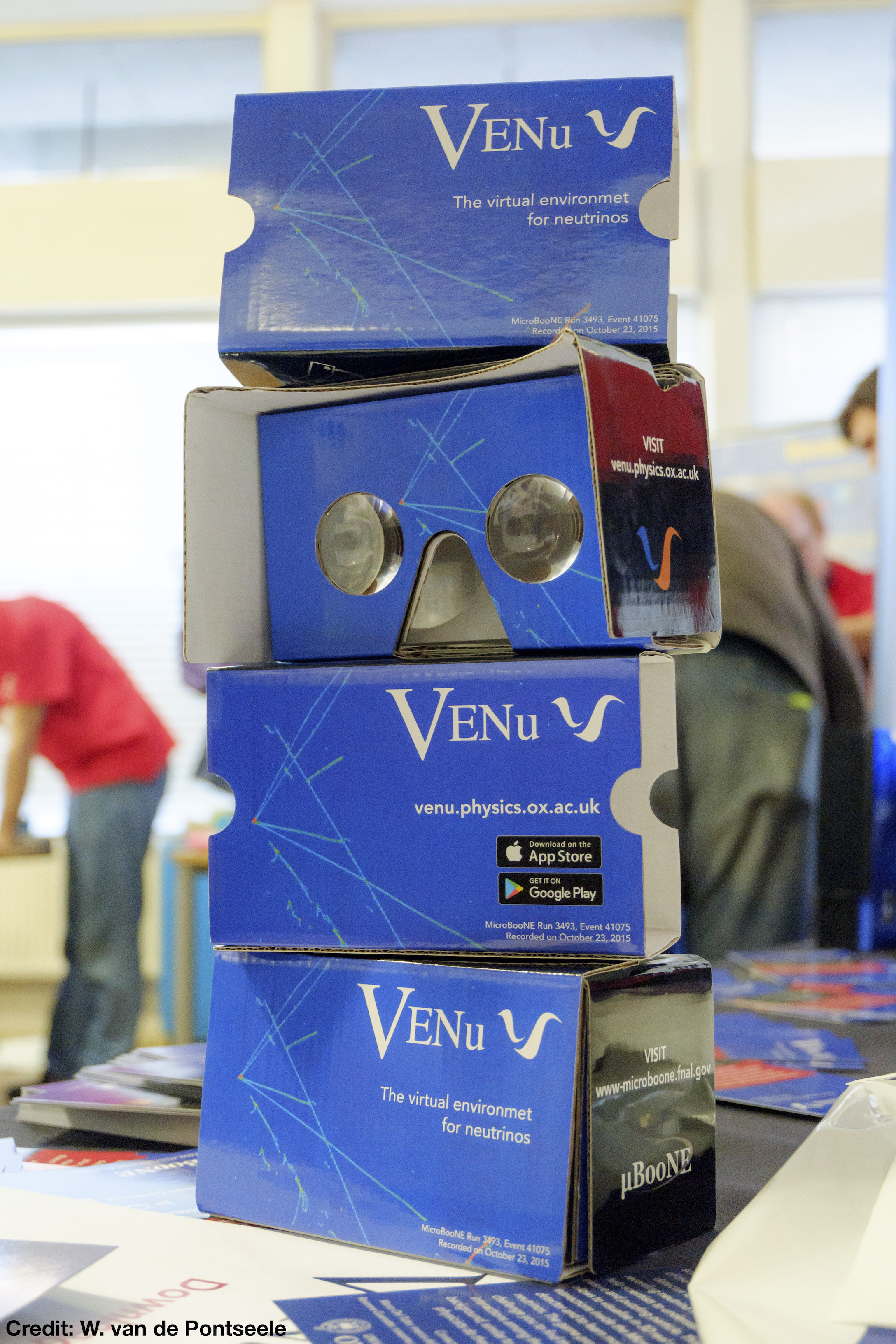}
   \label{fig:venu12}} \quad
\subfloat[][Experiencing the VENu Virtual Reality.]
   {\includegraphics[width=.373\textwidth]{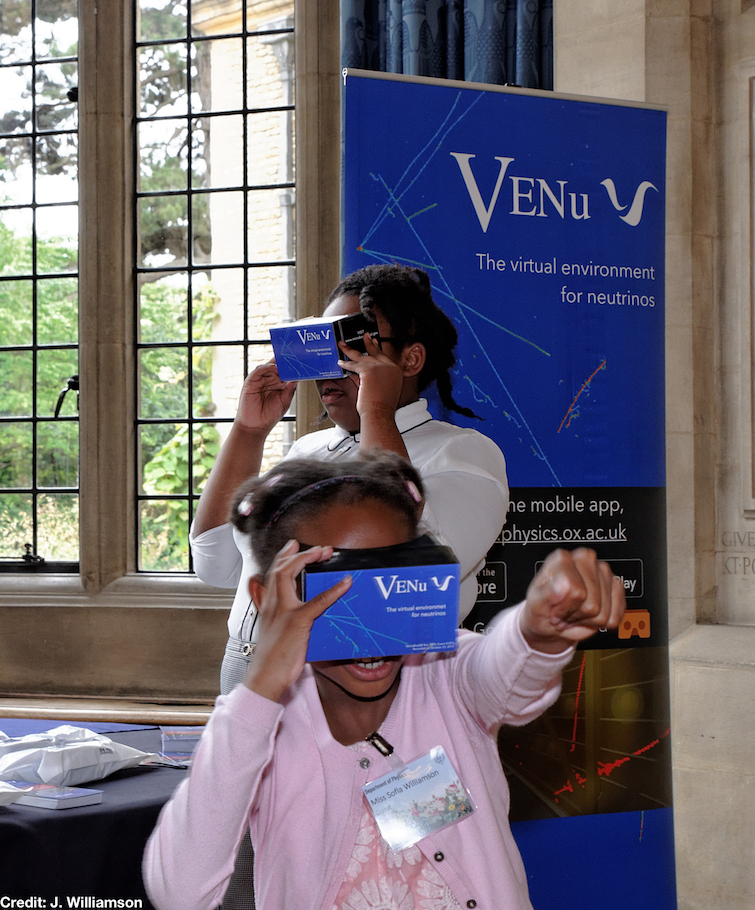}
   \label{fig:garden_party}} \\
\caption{}
\label{fig:stargazing}
%\end{adjustwidth}
\end{figure}

For the occasion, custom VR googles (Google Cardboards) were designed and provided to give to people coming to the event, so that they could try it at home, as can be seen in Figure \ref{fig:stargazing}. Google Cardboards are a powerful gadget to attract people during outreach events, and it turned out to be a very good way to promote the experiment and increase excitement in neutrino physics. 

We recorded that we gave away about 250 Cardboards. We then estimated the number of people reached during the launch event to be 750 people. We reached 300 downloads during the first three days from the launch. Apart from outreach events, the number of downloads is the most robust metric to understand how many people were reached.

People tried the app at the VENu stall during the Stargazing event and their impressions and feedback were monitored allowing us to understand what could be improved or added to the app.

%+++++++++FIGURE+++++++++
%++++++++++++++++++++++++

\section{Impact}

After the launch we were contacted by the \emph{International Business Time}, the Oxford local TV \emph{That's Oxfordshire}, the \emph{Institute of Physics} and \emph{Fermilab} to release interviews about our project. Interviews and podcasts can be found at these references: \cite{ibt}, \cite{thatsox}, \cite{iop}, \cite{fermilab}.

At the time of the submission of these proceedings, the app has reached more than 2000 downloads. The app being written in English is currently mainly targeted to the United Kingdom and the United States, but it is available to download from all over the world. In addition to
the U.K. and U.S., we have downloads from Canada, Italy, China, Germany, Australia, Switzerland, France and India. 

\begin{figure}[t]
%\begin{adjustwidth}{-2cm}{-2cm}
\centering
\subfloat[][iOS installs. Different colours just shows different countries.]
   {\includegraphics[width=.47\textwidth]{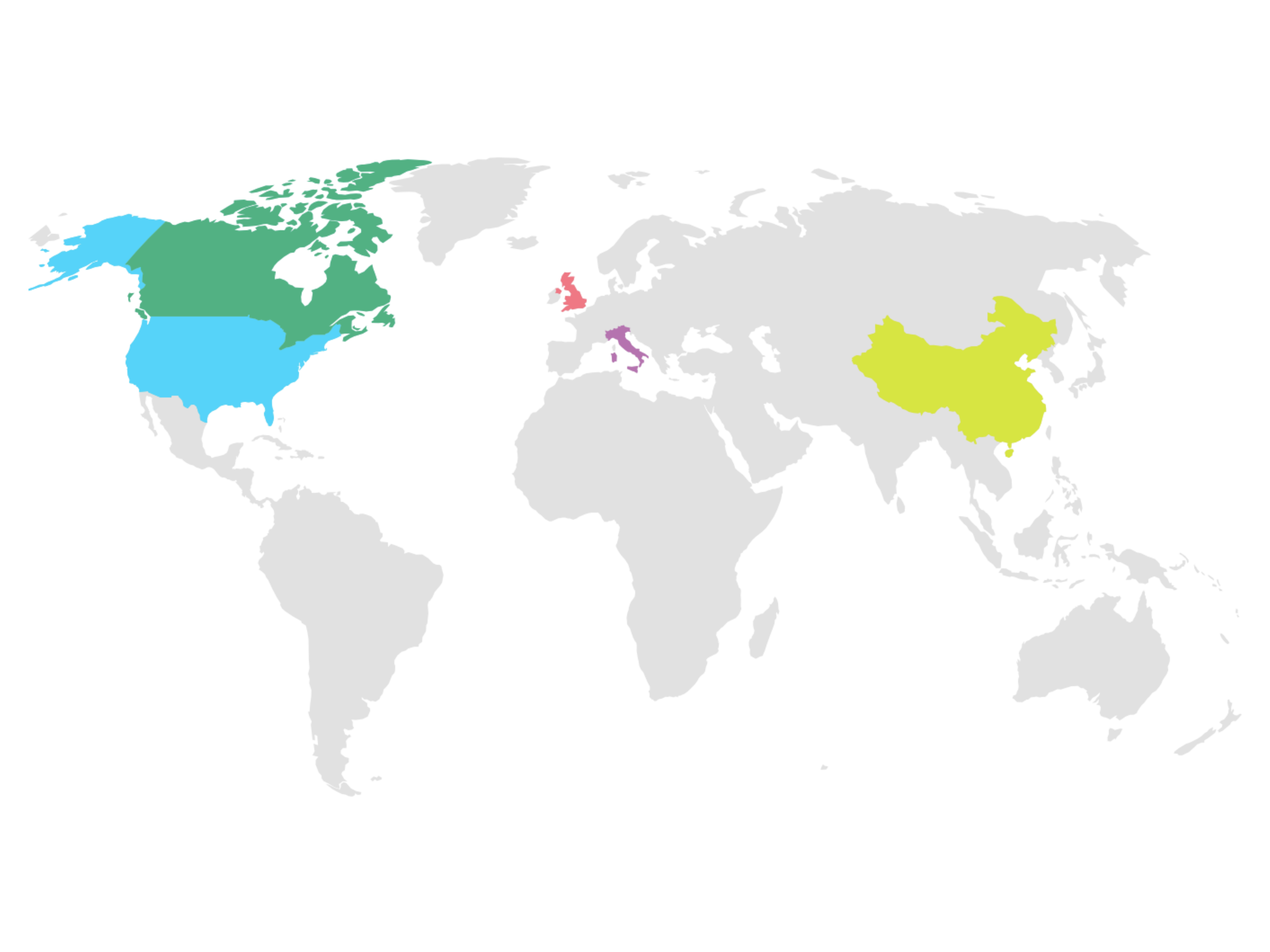}
   \label{fig:ioscountries}} \quad
\subfloat[][Android installs. The darker the blue the larger the number of download is.]
   {\includegraphics[width=.47\textwidth]{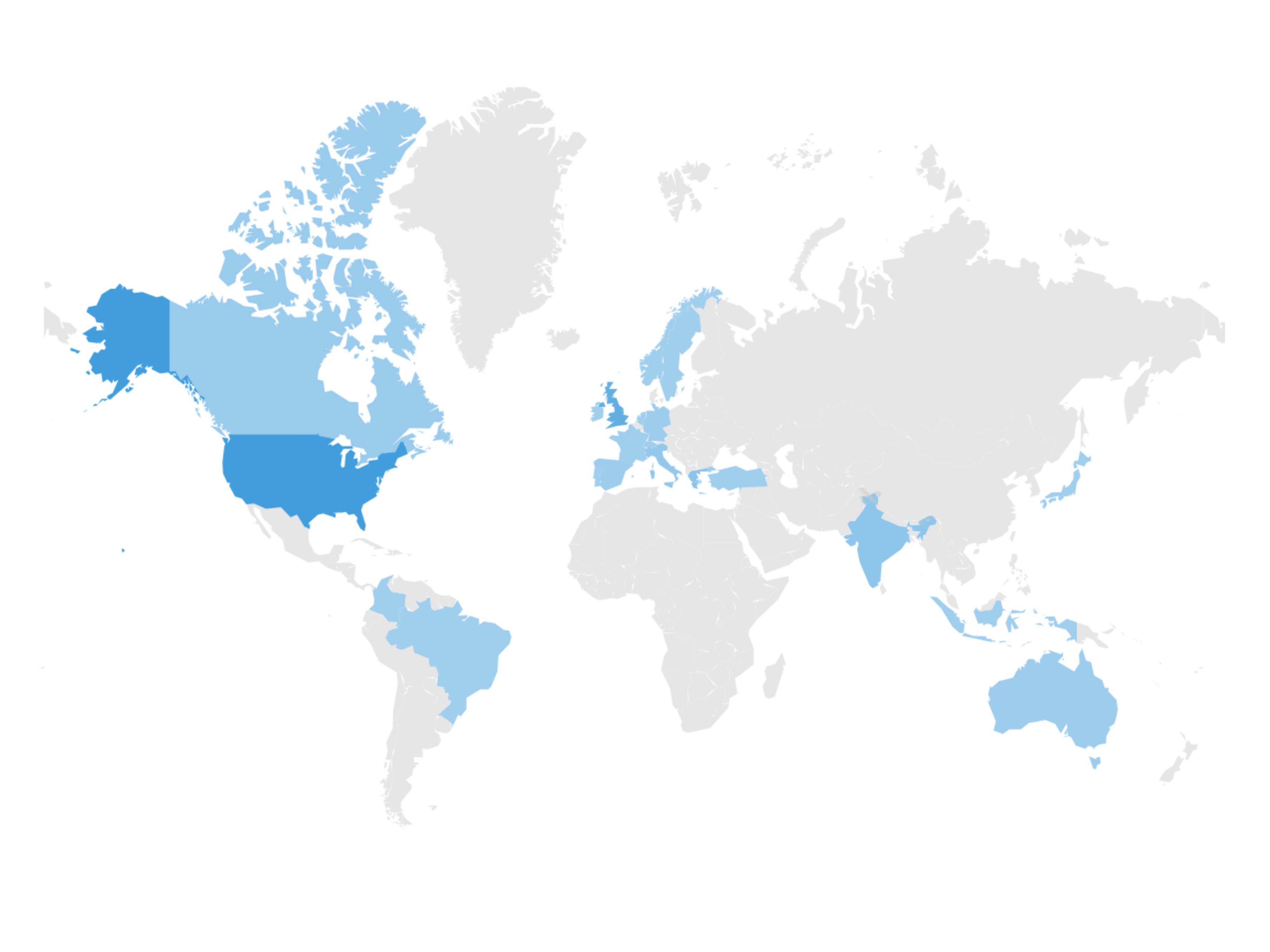}
   \label{fig:androidcountries}} \\
\caption{Locations of VENu installs by country in which the device was registered.}
\label{fig:countries}
%\end{adjustwidth}
\end{figure}

The iOS version of the app has been downloaded over 1850 times since released on January 28, 2017. The breakdown by country is shown in figure \ref{fig:ioscountries}.

The Android version of the app has amassed 701 downloads since release. By country in which the device was registered, this breaks down as shown in figure \ref{fig:androidcountries}.

%+++++++++FIGURE+++++++++
%++++++++++++++++++++++++

\section{Conclusions}

VENu is clearly making a strong impact during outreach events. It's providing a powerful visual tool with just a smartphone and a VR headset. We found that the public is really interested to learn about physics by experiencing what we are doing in 3D. This will be of great benefit to all future outreach events and is currently used at the Fermi National Accelerator Laboratory and universities worldwide.  

The public now has a simple and accessible way to look into what physicists are doing. This is as easy as downloading an app on their smartphone. Although the app is targeted to school aged users, having it on the Apple and Android marketplaces makes it available to a much larger audience. Everyone with an interest in science can check out the app from the online stores. To monitor people's impressions, the app features a feedback section were the users can email the VENu team with their thoughts. They can also leave reviews on the Apple and Google Stores. People satisfaction is demonstrated by the comments received and by the rating average that now amounts to 5/5 on the Apple Store and 4.8/5 on the Google Store.

\renewcommand*{\bibfont}{\footnotesize}
%\nocite{*}
\printbibliography
\begin{comment}

\end{comment}

\end{document}